\documentclass[12pt]{article}
\usepackage{amsmath,amssymb,epsfig,cite}
\usepackage{color}

\makeatletter

\@addtoreset{equation}{section} \makeatother

\begin{document}

\titlepage
\begin{flushright}
 \small SU-4252-834
\end{flushright}
\vskip 1cm
\begin{center}
{\large \bf Light Propagation in a Background Field for Time-Space
Noncommutativity and Axionic Noncommutative QED}
\end{center}

\vspace*{5mm} \noindent

\centerline{N.~Chatillon\footnote{nchatill@physics.syr.edu} and
A.~Pinzul\footnote{apinzul@physics.syr.edu}}

\centerline{ \em Department of Physics, Syracuse University,
Syracuse, NY 13244-1130, USA.} \vskip 0.5cm

\begin{center}
{\bf Abstract}
\end{center}

We study the low-energy effects of space-time non-commutativity on
light propagation in a background electromagnetic field. Contrary
to some of the previous claims, we find no polarization rotation
for vanishing time-space commutator $[\hat{x}^i,\hat{x}^0]= 0$,
although dispersion relation is modified, allowing for propagation
faster than the vacuum speed of light. For non-zero
$[\hat{x}^i,\hat{x}^0]$, as allowed with a proper quantization, a
naive rotation effect is found to be actually absent when physical
fields are defined through Seiberg-Witten map. We also consider
non-commutative QED weakly coupled to small mass particles such as
axions. Non-commutativity is found to dominate the inverse
oscillation length, compared to axion mass and QED effects, for
mixing particle masses smaller than $10^{-12}\ eV$. Conventional
constraints on axion coupling based on photon-axion transition
rates are unmodified, however induced ellipticity is proportional
to the non-commutativity squared length scale. This last effect is
found to be too small to account for the ellipticity reported by
the PVLAS experiment, yet unexplained by conventional QED or axion
physics.

\vskip 1cm

\section{Introduction}

The idea that space-time may be described by non-commutative (NC)
coordinates at short distance scales has attracted an increasing
interest. The expectation that, in a quantum gravity framework,
coordinates should obey an uncertainty relation
\cite{Doplicher:1994zv} of the form $\Delta \hat x^\mu \Delta \hat
x^\nu \geq \frac{1}{2}|\theta^{\mu\nu}|$, leads to the commutation
relations
\begin{equation}
[\hat x^\mu,\hat x^\nu]=i\theta^{\mu\nu}
\end{equation}
with $\theta^{\mu\nu}$ an antisymmetric set of constants. A
similar result is obtained in string theory in a non-vanishing
background of the NS-NS 2-form $B_{\mu\nu}$
\cite{Chu:1998qz,Schomerus:1999ug,Seiberg:1999vs}. This algebra of
abstract coordinates $\hat x^\mu$ can be realized on functions of
the commutative space-time coordinates, $x^\mu$, by replacing all
products with a star product $\star$, such that
\begin{equation}
[x^\mu,x^\nu]_\star = x^\mu \star x^\nu - x^\nu \star x^\mu = i
\theta^{\mu\nu} = [\hat x^\mu,\hat x^\nu].
\end{equation}
The specific choice of the Moyal product
\begin{equation}
f \star g = f\ \exp(\frac{i}{2}\overleftarrow{\partial_\mu}
\theta^{\mu\nu}\overrightarrow{\partial_\nu})\ g = f g +
\sum_{n=1}^\infty \frac{1}{n!}(\frac{i}{2})^n \theta^{\mu_1\nu_1}
... \theta^{\mu_n \nu_n}(\partial_{\mu_1}...\partial_{\mu_n}
f)(\partial_{\nu_1}...\partial_{\nu_n} g) .
\end{equation}
corresponds to a symmetric ordering prescription when mapping a
function of the commutative $x^\mu$ to an abstract function of the
NC $\hat x^\mu$, e.g. $x^\mu x^\nu \rightarrow \frac{1}{2}(\hat
x^\mu \hat x^\nu + \hat x^\nu \hat x^\mu)$. This product has the
important properties of breaking Lorentz invariance  due to the
presence of the constant tensor $\theta^{\mu\nu}$, and introducing
non-locality due to the infinite number of derivatives involved. A
twisted form of Lorentz invariance is however formally preserved
\cite{Chaichian:2004za}.

Following this approach, non-commutative QED had been built
\cite{Hayakawa:1999yt}, as well as a non-commutative Standard
Model \cite{Chaichian:2001py}. Its tree-level structure is
identical to the commutative SM, plus $O(\theta)$ corrections of
dimension 6 and higher. The strongest bounds on the NC energy
scale $\theta^{-1/2}$ come from the low energy constraints on
Lorentz violation. Assuming that $\theta^{\mu\nu}$ is constant in
the CMB frame, its orientation varies in an Earth based laboratory
frame, essentially due to Earth rotation. The spacelike components
$\theta^{ij}$ couple to atomic and nuclear angular momenta
similarly to a magnetic field, while the bounds on the timelike
components $\theta^{0i}$ are usually much weaker, if considered at
all. Clock comparaison experiments searching for a sidereal time
variation of the atomic Zeeman splitting yield a lower bound on
$\theta^{-1/2}$, that we will refer to as the spacelike component
scale in what follows. Tree-level analysis of the effect of
dimension 6 operators in NC QED implies $\theta^{-1/2}
> O(10 \ TeV)$, or $O(10\ GeV)$ for the $\theta^{0i}$ components \cite{Carroll:2001ws}. An estimation
of  the tree-level NC QCD sector gives the much more stringent
bound $\theta^{-1/2} > 5.10^{14}\ GeV$ \cite{Mocioiu:2000ip}, from
generated operators of the form $\theta^{\mu\nu}\bar N
\gamma_{\mu\nu} N$ involving nucleons.

Lorentz violating loop-induced operators of dimension 4 and lower
have also been considered. In a Lorentz-violating theory, these
very constrained operators are expected to be naturally large, and
grow with the UV momentum cut-off $\Lambda$, although
non-intuitive behavior also appears from the UV/IR connection. In
the NC QED sector, the constraint $\theta^{-1/2}
> 10^{9-10}\Lambda$ is obtained from the operator $m_e
\theta^{\mu\nu}\bar \psi \gamma_{\mu\nu}\psi$
\cite{Anisimov:2001zc}, and the stronger one $\theta^{-1/2}>
10^{13}\Lambda$ in the case of a supersymmetric regularization
where $\Lambda$ is the superpartner mass scale
\cite{Carlson:2002zb}. A similar analysis in the NC QCD sector
produces the even stronger $\theta^{-1/2} > 10^{14-15}\Lambda$
from clock comparison experiments \cite{Carlson:2001sw}. Thus,
even for a UV cut-off as small as $TeV$, the worst bounds push
$\theta^{-1/2}$ to the Planck scale. Also, the large dimensionless
ratio of the NC scale to the cut-off may be considered as
unnatural.

Taking this into account, finding new signals of non-commutativity
is certainly extremely challenging. In the best case, assuming
that NC does not apply at all to the QCD sector, and that
loop-induced operators are cancelled by a fine-tuning of the
counterterms, one may retain $\theta^{-1/2} > 10\ TeV$.

It appears that some of the most constraining tests of space-time
non-commutativity involve dynamics in a background
field\footnote{See also the recent work \cite{Marculescu:2006ca}
on wave solutions of the NC Einstein-Maxwell equations in absence
of a background field.}. It is interesting in particular to study
the propagation of electromagnetic waves in a non-zero background.
This was done in \cite{Cai:2001az} for the purely magnetic case
and \cite{Guralnik:2001ax} including an electric background too.
Both assumed $\theta^{0i}=0$, with the conclusion that the
dispersion relation is modified according to
\begin{equation}
\omega^2 = \vec{k}^2(1-2g\vec{\theta}_T\cdot \vec{\bf B}_T)
\end{equation}
to first order in $\theta$. We have defined the 3-vector $\theta^i
=\frac{1}{2}\epsilon^{ijk}\theta^{jk}$ \-; here the $T$ indices
refer to the projection transverse to the wave momentum. A first
important consequence is that depending on the sign of
$\vec{\theta}_T\cdot \vec{\bf B}_T$ the propagation may become
superluminal , i.e. the group velocity $\frac{d\omega}{dk}>1$. A
similar phenomenon has been found for NC solitons which may have
superluminal velocities on arbitrarily large distances
\cite{Bak:2000im,Hashimoto:2000ys}. It is discussed in
\cite{Hashimoto:2000ys} that this does not necessarily imply a
violation of causality because in that case the faster-than-light
propagation is associated to a preferred frame where
$\theta^{0i}=0$.

It has been argued \cite{Gomis:2000zz} that unitarity is violated
in NC theories unless there exist a frame where $\theta^{0i}=0$,
or frame independently,
\begin{eqnarray}
&\theta_{\mu\nu}\theta^{\mu\nu}=
\theta^{ij}\theta^{ij}-2\theta^{0i}\theta^{0i}>0&
\nonumber \\
{\rm and \ \
}&\epsilon_{\mu\nu\rho\sigma}\theta^{\mu\nu}\theta^{\rho\sigma}=0
,& \label{unitarity constraint}
\end{eqnarray}
which has lead many authors to ignore the $\theta^{0i}$
components. We stress that even in the case when there is such a
frame, it may be different from the laboratory frame, although
taking $\theta^{0i}=0$ e.g. in the CMB frame would lead to only a
very small contribution to $\theta^{0i}$ in the Earth frame, with
$\beta \sim 10^{-3}$. Also, unitarity no longer requires
(\ref{unitarity constraint}) if a proper quantization is applied
\cite{Bahns:2002vm}. As a result, the theories with
$\theta^{0i}\ne 0$ have become an area of active research (see
\cite{Balachandran:2004rq,Balachandran:2004yh,Balachandran:2004cr,Vassilevich:2004sj,Saha:2006qt}
for a non-exhaustive list of related works dealing with different
aspects of time-space noncommutativity). Therefore it is natural
that the properties of wave propagation with non-zero
$\theta^{0i}$ should be considered too. This was studied in
\cite{Abe:2003ys,Zahn:2004gt} for a magnetic background.

The recent claim, by the PVLAS collaboration, of having observed a
non-zero rotation and ellipticity of initially polarized light
propagating in a strong magnetic field \cite{Zavattini:2005tm}
provides another motivation for reconsidering NC wave propagation
with non-vanishing $\theta^{0i}$. Such effects typically result
from self-interaction terms of the electromagnetic field. They are
expected in ordinary QED from the loop-level photon splitting and
vacuum polarization graphs in a background field
\cite{Iacopini:1979ci}. The experiment was actually designed to
observe this last effect, but obtained instead larger results
unexplainable in the Standard Model, in particular an ellipticity
signal four orders of magnitude larger than the expected QED one.
There was a claim \cite{Chaichian:2005gh}, that the pure NC QED in
the case of the space-space non-commutativity ($\theta^{0i}=0$)
might be (partially) responsible for this effect. In this paper,
we revisit this proposal and arrive at a different conclusion,
namely that there is no rotation effect due to NC. In section 2,
we present the analysis of the general case including
$\theta^{0i}\neq 0$ and find that using the NC gauge field
variable, a qualitatively new effect seems to appear, with a
polarization rotation. However this effect is an artifact of using
the transverse gauge inconsistently for NC gauge field variables,
which would involve an infinite number of time derivatives. We
show that the effect is still vanishing when using physical gauge
fields obtained via the Seiberg-Witten map.

Another possible source of polarization rotation and ellipticity
generation is the mixing of photons with a light scalar or
pseudo-scalar such as an axion \cite{Maiani:1986md}. The PVLAS
results could be explained by an axion mass $m\sim 10^{-3}\ eV$
and a photon-axion coupling scale $M^{-1}\sim (10^6 \ GeV)^{-1}$,
however this last value would induce a globular cluster star
cooling rate \cite{Raffelt:1996wa}, and a solar axion flux in the
CAST detector \cite{Zioutas:2004hi} by Primakoff photon-axion
conversion, by four orders of magnitude too large. Astrophysical
arguments \cite{Masso:2005ym,Jain:2005nh,Jaeckel:2006id} have been
proposed recently to evade the previous bounds, along with simple
alternatives to axion models based on the mixing with extra vector
fields \cite{Masso:2006gc,Antoniadis:2006wp}. In section 3, we
study wave propagation in a background field in axionic NC QED. We
show that the axion-photon inverse oscillation length, which is
the relevant parameter for induced ellipticity, can be dominated
by non-commutativity for small enough axion masses, and still be
larger than the ordinary QED contribution. The axion coupling $M$
needed is however too small to reproduce the ellipticity observed
in the specific PVLAS setup. Finally, we discuss the constraints
on the various scales relevant for the problem.


\section{Wave propagation in a background field for
general non-commutativity}

\subsection{Noncommutative vs. physical gauge field variables}

We consider non-commutative QED, characterized by the constants
$i\theta^{\mu\nu}=[\hat x^\mu, \hat x^\nu]$. Alternatively, one
may use the magnetic part $\theta^i
=\frac{1}{2}\epsilon^{ijk}\theta^{jk}$ and the electric part
$\xi^i=\theta^{0i}$. The action is obtained by substitution of
Moyal products $\star$ to ordinary products :
\begin{eqnarray}
{\cal S}_{NC\ QED} &=& \int d^4x \Big(-\frac{1}{4}{\cal
F}^{\mu\nu}\star {\cal F}_{\mu\nu}\Big)
\nonumber \\
{\cal F}^{\mu\nu} &=& \partial^{[\mu} {\cal A}^{\nu]}+ig[{\cal
A}^\mu,{\cal A}^\nu]_\star\ \approx \partial^{[\mu} {\cal
A}^{\nu]} -g \theta^{\alpha\beta}\partial_\alpha {\cal
A}^\mu\partial_\beta {\cal A}^\nu + O(\theta^2)
\end{eqnarray}
with $g$ the $U(1)$ gauge coupling. It can be seen that even
abelian gauge theories possess photon self-couplings, suppressed
by the NC energy scale. A single $\star$ may be removed from every
term in the action, up to a total derivative. This action does not
enjoy the usual $U(1)$ gauge invariance, as can be seen in
particular from the fact that it does not depend only on the
"commutative" field-strength $\partial_{[\mu} {\cal A}_{\nu]}$ .
The full NC gauge invariance is expressed as
\begin{eqnarray}
{\cal A}_\mu &\rightarrow& U^\dagger \star ({\cal A}_\mu
+\frac{1}{ig}\partial_\mu)\star U
\nonumber \\
(U\star U^\dagger)(x)&=& (U^\dagger\star U)(x)= \bf 1 .
\end{eqnarray}
This can be traced back to the fact that the (Moyal) commutator of
two infinitesimal gauge transformations involves now the
commutator of functions :
\begin{equation}
[\epsilon_1^a(x)T_a,\epsilon_2^b(x)T_b]_\star
=\frac{1}{2}\{\epsilon_1^a(x), \epsilon_2^b(x)\}_\star\ [T_a,T_b]+
\frac{1}{2}[\epsilon_1^a(x),\epsilon_2^b(x)]_\star\ \{T_a,T_b\}
\end{equation}
and is thus no longer vanishing even in the $U(1)_{NC}$ case where
$[T_a,T_b]=0$. The standard trick to recover the ordinary gauge
symmetry is to perform a nonlocal redefinition of the gauge (and
matter) fields according to the Seiberg-Witten map
\cite{Seiberg:1999vs}:
\begin{eqnarray}
{\cal A}_\mu &=& A_\mu
-\frac{1}{2}g\theta^{\alpha\beta}A_\alpha(\partial_\beta A_\mu +
F_{\beta\mu})+O(\theta^2)
\nonumber \\
F_{\mu\nu}&=&\partial_{[\mu}A_{\nu ]}\ ,
\end{eqnarray}
which is unique up to some homogeneous terms that for our purpose
can be consistently set to zero \cite{Pinzul:2004tq}. The field
$A_\mu$ is the commutative $U(1)$ gauge field and the gauge
invariant action is
\begin{equation}
{\cal S}_{NC\ QED}= \int d^4x
\Big(-\frac{1}{4}F^{\mu\nu}F_{\mu\nu}+\frac{1}{8}g\theta^{\alpha
\beta}F_{\alpha\beta}F^{\mu\nu}
F_{\mu\nu}-\frac{1}{2}g\theta^{\alpha\beta}F_{\mu\alpha}F_{\nu\beta}
F^{\mu\nu}\Big) + O(\theta^2) . \label{NC QED action in
commutative fields}
\end{equation}
It is interesting to compare this approach to NC QED with the one
in which calculations are done working directly with the
non-commutative gauge field $\mathcal{A}$ (this is the approach
used in \cite{Chaichian:2005gh}). Though it enjoys an obvious
advantage of the possibility to perform calculation exact in
$\theta$, in the case of the time-space non-commutativity,
$\theta^{0i}\neq 0$, this approach meets serious difficulties
related to the gauge fixing (when $\theta^{0i}= 0$, this problem
is absent). The `natural' gauge used in \cite{Chaichian:2005gh} -
the temporal transverse gauge ${\cal A}_0 =
\partial_i {\cal A}^i=0$ - is now inconsistent with the equations of motion,
for $\vec{\xi} \neq 0$. The equations of motion for the field
perturbation read, prior to gauge fixing, and \emph{to all order
in $\theta^{\mu\nu}$} for constant background field strength:
\begin{eqnarray}
&\Big[\Box \eta^{\mu\nu}-\partial^\mu \partial^\nu -
g\theta^{\rho\sigma}(\partial_\rho {\bf\cal
A}_B^\tau)(2\eta^{\mu\nu}\partial_\tau-\partial^\mu
\delta^\nu_\tau-\partial^\nu\delta^\mu_\tau)\partial_\sigma&
\nonumber \\
&+g^2(\theta^{\alpha\beta}\partial_\alpha {\cal A}_B^\tau)
(\theta^{\rho\sigma}\mathcal{A}_B^\tau)\eta^{\mu\nu}\partial_\beta
\partial_\sigma-g^2(\theta^{\alpha\beta}\partial_\alpha {\cal A}^\mu_B)
(\theta^{\rho\sigma}\partial_\sigma {\cal A}_B^\nu)
\partial_\beta \partial_\sigma \Big] {\cal A}_\mu + O({\cal A}^2)=0&
\end{eqnarray}
where ${\bf\cal A}^\mu_B$ is the background vector potential. From
this the transverse temporal gauge would impose
\begin{equation}
(\vec {\bf \cal B}\times \vec{\xi})\cdot \partial_0^2 \vec{\bf
\mathcal{A}}=0\ ,
\end{equation}
which would kill one of the two required degrees of freedom. So
the naive `natural' gauge choice is inconsistent.\footnote{This is
important : using this gauge, one would have arrived to the
conclusion that when $\theta^{0i}\neq 0$ there is rotation of the
polarization. Which is not the case as we will see in the next
section.} Therefore we choose to work instead using (\ref{NC QED
action in commutative fields}) with commutative gauge field
variables obtained from the Seiberg-Witten map, at the cost of
being limited to work at finite order in $\theta^{\mu\nu}$.

\subsection{Explicit derivation of the dispersion relation for
$\theta^{0i}\neq 0$\label{theta0i section}}

We consider now the quadratic expansion of the action (\ref{NC QED
action in commutative fields}) around a constant non-zero
background, $F_{\mu\nu}={\bf F}_{\mu\nu}+f_{\mu\nu}$ with $f \ll
{\bf F}$. It turns out to be convenient to work with explicitly
gauge invariant variables $E^i$ and $B^k$, which are defined as in
the commutative case
\begin{equation}\label{EB}
F^{i0}=E^i\ ,\ F_{ij}=-\epsilon_{ijk}B^k\ .
\end{equation}
As above, we will use two 3-d vectors, $\xi^i$ and $\theta^k$,
characterizing non-commutativity:
\begin{equation}\label{ncpar}
\theta^{0i}=\xi^i\ , \ \theta^{ij}=\epsilon^{ijk}\theta^k\ .
\end{equation}
So we explicitly take into account the possibility of time-space
non-commutativity\footnote{After completion of this work,
reference \cite{Abe:2003ys} which considered the dispersion
relations in the $\xi^i \neq 0$ case for a constant magnetic
background, purpose of the section \ref{theta0i section}, was
brought to our attention. The work \cite{Zahn:2004gt} showed later
that the same result could be found using covariant coordinates.}.
In the case when $\xi^i = 0$, our analysis reduces to the one
presented in \cite{Guralnik:2001ax}.

In terms of the defintions (\ref{EB}) and (\ref{ncpar}) the
Lagrangian from Eq.(\ref{NC QED action in commutative fields})
becomes
\begin{equation}\label{NCL}
 L =
\frac{1}{2}(1+g\vec{\xi}\cdot\vec{E}-g\vec{\theta}\cdot\vec{B})(\vec{E}^2
- \vec{B}^2) +
(g\vec{\xi}\cdot\vec{B}+g\vec{\theta}\cdot\vec{E})(\vec{E}\cdot\vec{B})\
.
\end{equation}

Varying (\ref{NC QED action in commutative fields}) with respect
to $A_i$ and $A_0$ gives, as usual, equation of motion and a Gauss
law constraint respectively
\begin{equation}\label{eom}
\frac{\delta S}{\delta A_i} = -\partial_0\left(\frac{\partial
L}{\partial
E^i}\right)+\epsilon^{ijk}\partial_k\left(\frac{\partial
L}{\partial B^j}\right)\ ,\ \frac{\delta S}{\delta A_0} =
\partial_i\left(\frac{\partial L}{\partial E^i}\right)\ .
\end{equation}
To write (\ref{eom}) explicitly, it is convenient to introduce
'displacement field' $\vec{D}$ and 'magnetic field' $\vec{H}$
\begin{eqnarray}\label{DH}
\vec{D}=(1+g\vec{\xi}\cdot\vec{E}-g\vec{\theta}\cdot\vec{B})\vec{E}
+ (g\vec{\xi}\cdot\vec{B}+g\vec{\theta}\cdot\vec{E})\vec{B}
+\frac{1}{2}(\vec{E}^2 - \vec{B}^2)g\vec{\xi} +
(\vec{E}\cdot\vec{B})g\vec{\theta}\ ,\\
\vec{H}=(1+g\vec{\xi}\cdot\vec{E}-g\vec{\theta}\cdot\vec{B})\vec{B}
- (g\vec{\xi}\cdot\vec{B}+g\vec{\theta}\cdot\vec{E})\vec{E}
+\frac{1}{2}(\vec{E}^2 - \vec{B}^2)g\vec{\theta} -
(\vec{E}\cdot\vec{B})g\vec{\xi}\ .\nonumber
\end{eqnarray}
Then Eq.(\ref{eom}) will have the form of the sourceless Maxwell
equations
\begin{eqnarray}\label{eom1}
\partial_0 \vec{D}-\vec{\partial}\times \vec{H}=0\ &,&\ \partial_0 \vec{B}+\vec{\partial}\times
\vec{E}=0\\
\vec{\partial}\cdot\vec{D} = 0\ &,& \ \vec{\partial}\cdot\vec{B} =
0\ . \nonumber
\end{eqnarray}

We are interested in the plane-wave solution of (\ref{eom1}),
$\vec{E}=\vec{\mathcal{E}}(\omega t -\vec{k}\cdot \vec{r})$, in
the presence of the constant magnetic background
$\vec{\mathbf{B}}$. Then we have from (\ref{eom1})
\begin{equation}\label{planewave}
\vec{B}=\frac{\vec{k}}{\omega}\times\vec{\mathcal{E}} +
\vec{\mathbf{B}}\ ,\
\vec{D}=-\frac{\vec{k}}{\omega}\times\vec{{H}} + \vec{\mathbf{D}}\
.
\end{equation}
The non-vanishing constant background field $\vec{\mathbf{D}}$ is
needed to satisfy (\ref{eom1}) for a vanishing background,
$\vec{\mathbf{E}}=\vec{0}$.

We will not attempt to solve (\ref{DH})-(\ref{planewave}) exactly
in fields. We rather will consider the linearized system which is
enough for our purpose - finding the dispersion relation. Writing
$\vec{B}= \vec{\mathcal{B}}+\vec{\mathbf{B}}$ and $\vec{H}=
\vec{\mathcal{H}}+\vec{\mathbf{H}}$, we have the following
linearized system
\begin{eqnarray}\label{linear}
{\mathcal{D}}_i = \varepsilon^{ij}{\mathcal{E}}^j +
\rho^{ij}{\mathcal{B}}^j \\
{\mathcal{H}}_i = \mu^{ij}{\mathcal{E}}^j +
\eta^{ij}{\mathcal{B}}^j \ ,\nonumber
\end{eqnarray}
where
\begin{eqnarray}\label{coeff}
\varepsilon^{ij} = \delta^{ij}(1 - g\vec{\theta}\cdot
\vec{\mathbf{B}}) + g\theta^i {\mathbf{B}}^j + g\theta^j
{\mathbf{B}}^i\ ,\  \rho^{ij} = \delta^{ij}(g\vec{\xi}\cdot
\vec{\mathbf{B}}) + g\xi^i {\mathbf{B}}^j - g\xi^j {\mathbf{B}}^i\ ,\\
\mu^{ij} = \delta^{ij}(1 - g\vec{\theta}\cdot \vec{\mathbf{B}}) -
g\theta^i {\mathbf{B}}^j - g\theta^j {\mathbf{B}}^i\ ,\  \eta^{ij}
= -\delta^{ij}(g\vec{\xi}\cdot \vec{\mathbf{B}}) + g\xi^i
{\mathbf{B}}^j - g\xi^j {\mathbf{B}}^i\ .\nonumber
\end{eqnarray}

Let us choose the third direction along $\vec{k}$. Then we have
$\vec{k}\cdot \vec{\mathcal{E}}=0$.\footnote{It is true that
longitudinal component of $\vec{\mathcal{E}}$, ${\mathcal{E}}_3$,
is not zero. But because it is of the order of $\theta$, one
cannot catch it in the linearized approach. But this does not
affect the calculation of the dispersion relation up to order
$O(\theta^2)$.} Now using the (homogeneous part of)
Eq.(\ref{planewave}) in Eq.(\ref{linear}) after some algebra one
arrives at the following dispersion relation
\begin{equation}\label{dispersion}
\frac{\vec{k}^2}{\omega^2} = \Big( 1 + 2g\vec{\theta}_T\cdot
\vec{\mathbf{B}}_T - 2(g\vec{\xi}\times \vec{\mathbf{B}})_3 \Big)+
O(\theta^2)\ ,
\end{equation}
where the subscript $T$ means the transverse component of the
corresponding vector. The important property of
Eq.(\ref{dispersion}) is that RHS is actually independent of the
polarization of the plane wave. This means that the only effect of
the noncommutativity is the change of 'speed of light', the same
for both polarization. So there is no rotation of the polarization
of light in this approximation ($O(\theta^2)$) in contrast with
the conclusion of \cite{Chaichian:2005gh}.

\subsection{Calculation from the Lorentz-violating extended Standard Model}

These results may be cross-checked and generalized using the
framework of the Lorentz-violating extension of QED
\cite{Kostelecky:2002hh}, described by the renormalizable
Lagrangian
\begin{equation}
{\cal L}_{LV\ QED}=-\frac{1}{4}F_{\mu\nu}F^{\mu\nu}
-\frac{1}{4}(k_F)_{\mu\nu\rho\sigma}F^{\mu\nu}F^{\rho\sigma} +
\frac{1}{2}
(k_{AF})^{\mu}\epsilon_{\mu\nu\rho\sigma}A^{\nu}F^{\rho\sigma}
\end{equation}
where $k_F$ and $k_{AF}$ are Lorentz-breaking constant tensors,
and one can assume the same symmetries as the Riemann tensor for
$k_F$. In the case of non-commutative space-time with constant
background field ${\bf F}_{\mu\nu}$, one can treat $A_{\mu}$ in
the action above as the perturbation around this background. The
CPT-odd term $(k_{AF})^\mu=0$ vanishes and
\begin{eqnarray}
&&(k_{F})_{\mu\nu\rho\sigma} =\frac{1}{8}(T_{[\mu\nu][\rho\sigma]}
+ \mu\nu \leftrightarrow \rho\sigma) \label{non-commutative k_F}
\nonumber \\
&&T_{\mu\nu\rho\sigma}\equiv -\frac{1}{2}g\theta^{\alpha\beta}{\bf
F}_{\alpha\beta} \eta_{\mu\rho}\eta_{\nu\sigma}
-g\theta_{\mu\nu}{\bf F}_{\rho\sigma} + 4 g\theta_{\alpha\nu}{\bf
F}_\rho^{\ \alpha} \eta_{\mu\sigma}+2g\theta_{\nu\sigma}{\bf
F}_{\mu\rho} + O(\theta^2)
\end{eqnarray}
where we defined $M_{[\mu\nu]}\equiv M_{\mu\nu}-M_{\nu\mu}$. The
photon dispersion relation is modified, for the two polarizations,
as follows :
\begin{equation}
\omega_{\pm}=(1+\rho \pm \sigma)|\vec{k}|
\end{equation}
with
\begin{eqnarray}
\rho &\equiv& \frac{1}{2} \tilde{k}^\alpha_{\ \alpha}
\nonumber \\
\sigma^2 &\equiv&
\frac{1}{2}\tilde{k}_{\alpha\beta}\tilde{k}^{\alpha\beta} - \rho^2
\nonumber \\
\tilde{k}_{\mu\rho} &\equiv& (k_F)_{\mu\nu\rho\sigma}
\frac{k^\nu}{|\vec{k}|}\frac{k^\sigma}{|\vec{k}|} .
\end{eqnarray}
The polarization rotation arises from the speed difference
$2\sigma$ between the polarization orthogonal and parallel to the
background field, and is in general non-zero. However in the NC
case, we find $\sigma=0$ to first order in $\theta$, leading to no
rotation.

We obtain the overall change in the speed of light from
\begin{equation}\label{dispersion1}
\frac{d\omega}{dk}-1= \rho =  g\theta_{\alpha \mu} {\bf
F}^{\alpha}_{\ \nu} \frac{k^\mu}{|\vec{k}|}\frac{k^\nu}{|\vec{k}|}
+ O(\theta^2)
\end{equation}
where to leading order in $\theta$, the photon momentum $k^\mu$ on
the right-hand side can be taken to obey the unmodified dispersion
relation $k^2=0$. Again, this can take both signs and illustrates
superluminal propagation in NC field theories.

It is easy to see that (\ref{dispersion1}) is nothing but our
result (\ref{dispersion}), including also a constant background
electric field. Thus calculation in the framework of the
Lorentz-violating extension of QED confirms our conclusion and
contradicts the one of \cite{Chaichian:2005gh} where a non-zero
rotation was claimed in spite of a vanishing $\sigma$.

\section{Non-commutative axionic electrodynamics}

\subsection{Equations of motion}

Let us consider now the coupling of a pseudoscalar $\phi$ to
electromagnetism in the NC framework. Although we will refer to it
as an axion in what follows, we will not assume that it is a
Peccei-Quinn axion introduced to solve the strong CP problem
\cite{Peccei:1977hh,Peccei:1977ur,Weinberg:1977ma,Wilczek:1977pj},
and we will keep $M$ and $m$ below as independent parameters.
Similarly, we will assume that the matter couplings to the
pseudo-scalar are vanishing or small enough to evade existing
constraints that would be stronger than those from the pure
axion-gauge sector. The corresponding action in NC field variables
is
\begin{eqnarray}
{\cal S}_{axion} &=& \int d^4x \Big(\frac{1}{2}(\partial
\phi)^2-\frac{m^2}{2}\phi^2 + \frac{\phi}{8M}
\epsilon_{\mu\nu\rho\sigma}({\cal F^{\mu\nu}}\star {\cal
F^{\rho\sigma}})\Big)
\nonumber \\
{\cal F^{\mu\nu}}&=& \partial^{[\mu} {\cal A^{\nu]}} +ig[{\cal
A}^\mu,{\cal A^\nu}]_\star
\end{eqnarray}
where one star product has been removed in each term, which leaves
the action invariant up to a boundary term. Notice that for
constant $\phi$, the last term is still a topological invariant.
Applying the Seiberg-Witten map ${\cal A}[A]$ and neglecting
$O(\theta \times \frac{1}{M})$ terms, the action reduces to the
commutative one :
\begin{eqnarray}
{\cal S}_{axion} &=& \int d^4x \Big(\frac{1}{2}(\partial
\phi)^2-\frac{m^2}{2}\phi^2 + \frac{\phi}{8M}
\epsilon_{\mu\nu\rho\sigma}F^{\mu\nu}F^{\rho\sigma}\Big) +
O(\frac{\theta}{M})
\nonumber \\
F^{\mu\nu}&=& \partial^{[\mu} A^{\nu]} .
\end{eqnarray}
We expand around a constant background $F_{\mu\nu}={\bf
F}_{\mu\nu}+f_{\mu\nu}$, and we assume $\tilde{\bf F}_{\mu\nu}{\bf
F}^{\mu\nu}=0$. The total action is
\begin{eqnarray}
{\cal S}&=&\int d^4x \Big(
-\frac{1}{4}[\eta_{\mu\rho}\eta_{\nu\sigma}+(k_F)_{\mu\nu\rho\sigma}]
f^{\mu\nu}f^{\rho\sigma}+\frac{1}{2}(\partial \phi)^2
-\frac{m^2}{2}\phi^2+\frac{1}{2M}\tilde{\bf
F}_{\mu\nu}f^{\mu\nu}\phi\Big)
\nonumber \\
&&+ O\Big(\theta^2,\frac{\theta}{M},f^3\Big)
\end{eqnarray}
with $k_F$ given in (\ref{non-commutative k_F}), and $\tilde{\bf
F}_{\mu\nu}\equiv \frac{1}{2}\epsilon_{\mu\nu\rho\sigma}{\bf
F}^{\rho\sigma}$ is the dual background electromagnetic field
strength.

We specialize now to the PVLAS-like configurations, with purely
magnetic background orthogonal to the wave propagation, and take
for simplicity pure space-space non-commutativity $\theta^{0i}=0$,
$\theta^i=\frac{1}{2}\epsilon^{ijk}\theta^{jk}$. The equations of
motion are, to first order in $\theta$ :
\begin{eqnarray}
E_L &=& O(\theta)
\nonumber \\
\Big(1-\frac{\vec{k}^2}{\omega^2}+2 g\vec{\theta}\cdot \vec{\bf
B}\Big)E_{T\bot}&=& 0
\nonumber \\
\Big(1-\frac{\vec{k}^2}{\omega^2}+2g \vec{\theta}\cdot \vec{\bf
B}\Big)E_{T\|} -\frac{|\vec{\bf B}|}{M} \phi&=& 0
\nonumber \\
(\omega^2-\vec{k}^2 -  m^2)\phi -\frac{|\vec{\bf B}|}{M}
E_{T\|}&=&0 .
\end{eqnarray}
Here $E_L$ is the longitudinal (to $\vec{k}$) component of the
electric field, $E_{T\bot}$ is the transverse component orthogonal
to the background magnetic field, and $E_{T\|}$ is the transverse
component projected on the background magnetic field ; these three
fields form an orthogonal basis. The longitudinal component,
non-vanishing here, is a function of the transverse ones and does
not represent a third degree of freedom for the electromagnetic
field.

The dispersion relation for the transverse component orthogonal to
$\vec{\bf {B}}$, $E_{T\bot}$, is the same as in the case of the
pure NC QED
\begin{equation}
\omega^2_{\bot}= \Big[1-2g(\vec{\theta}\cdot \vec{\bf
B})\Big]\vec{k}^2 \ .\label{omega_bot}
\end{equation}
Searching for plane wave solutions for the $\phi$-$E_{T\|}$
coupled system, we obtain the energy eigenvalues
\begin{equation}
(\omega^\pm_{\|})^2=\frac{1}{2}\Big(\omega_\bot^2+\omega_\phi^2+\frac{\vec{\bf
B}^2}{\tilde M^2}\Big)\pm
\frac{1}{2}\sqrt{\Big(\omega_\bot^2+\omega_\phi^2+\frac{\vec{\bf
B}^2}{\tilde M^2}\Big)^2-4 \omega_\bot^2 \omega_\phi^2}
\label{omega_parallel}
\end{equation}
where we introduced the notation
\begin{equation}
\omega^2_\phi\equiv \vec{k}^2 + m^2
\end{equation}
which is the axion squared energy in the absence of mixing, and
$\tilde{M}\equiv M(1+g\vec{\theta}\cdot \vec{\bf B})\approx M$ to
lowest order. The effect of non-commutativity is thus the
modification (\ref{omega_bot}) of the photon dispersion relation.
In the $\theta\rightarrow 0$ limit for finite $M$, the usual
commutative axion-photon mixing occurs \cite{Maiani:1986md}.

In the $M\rightarrow \infty$ limit for finite $\theta$, the axion
and parallel photon decouple, leading to the known
\begin{eqnarray}
\omega^2_{\rm axion} &=& \omega^2_\phi = \vec{k}^2 +m^2
\nonumber \\
\omega^2_{\rm photon} &=& \omega_\bot^2=
\Big[1-2g(\vec{\theta}\cdot \vec{\bf B})\Big]\vec{k}^2
\end{eqnarray}
where speed of light is equally shifted for both polarizations.

\subsection{Physical scales and constraints}

Let us consider now the scales involved in the problem. The beam
pulsation corresponds to infrared radiation $\omega\sim 1.2\ eV$.
We assume that axions are relativistic, $m^2 \ll \vec{k}^2\sim
(1.2\ eV)^2$. These conditions exclude the detection of resonant
axion production stimulated by an hypothetical galactic halo dark
matter axion background \cite{Sikivie:1983ip}, which would require
$\frac{\omega}{m}-1\leq \frac{1}{2}(\frac{v}{c})_{DM}^2\sim
10^{-6}$. The background magnetic field is of order $|\vec{\bf
B}|\sim 5.5\ T \Rightarrow g |\vec{\bf B}| \sim (18\ eV)^2$. Weak
axion coupling such that $\vec{\bf B}^2/M^2 \ll \omega^2$ is
assumed, implying $gM \gg 15\ eV$, obviously satisfied when
considering other experimental constraints on $M$.

Let us remind the allowed ranges of the axion parameters $M,m$.
The strongest \emph{model-independent} constraints on $M$, which
make no assumption on the axion matter couplings and keep $M,m$ as
independent parameters, come from the Primakoff photon to axion
conversion in stars. Avoiding excessive star cooling by this
process in the globular clusters, which would modify the star
evolution, imposes \cite{Raffelt:1996wa}
\begin{equation}
M \geq 1.7\times 10^{10}\ GeV . \label{star primakoff constraint
on M}
\end{equation}
A bound of the same order is obtained from the non-observation of
an axion flux from the Sun by the CAST helioscope
\cite{Zioutas:2004hi}, which reconverts them into photons in a
strong magnetic field :
\begin{equation}
M \geq 0.86\times 10^{10}\ GeV .
\end{equation}
More specifically now, the \emph{QCD axion} required to solve the
strong CP problem has matter couplings almost completely
characterized by the Peccei-Quinn symmetry breaking scale $f$,
related to its photon coupling $M^{-1}$ by $ f = g_\gamma
\frac{g^2}{4\pi^2}M$ with $g_\gamma$ a model-dependent coefficient
of order unity. Its mass $m$ is given by
\begin{equation}
m= (0.62\times 10^{-3}\ eV)\ \frac{10^{10}\ GeV}{f}\sim 2.7
g_\gamma^{-1}\ 10^{-3}\ eV \frac{10^{10}\ GeV}{M}\ .
\end{equation}
Furthermore, its Yukawa coupling to nucleons $g_{\phi N}\sim
\frac{1\ GeV}{f}$ is constrained to be $g_{\phi N}\leq 3\times
10^{-7}$ to avoid excessive supernovae cooling
\cite{Ellis:1987pk}, which pushes higher the bound (\ref{star
primakoff constraint on M}) by roughly one order of magnitude.
Correspondingly, this translates into
\begin{equation}
m\leq 10^{-2}\ eV
\end{equation}
for the QCD axion mass. A lower bound
\begin{equation}
m \geq 10^{-6}\ eV \label{cosmo bound on axion mass}
\end{equation}
can be set by requiring that the QCD axions, produced by the
vacuum misalignment mechanism \cite{Preskill:1982cy} or by axionic
string radiation \cite{Battye:1993jv}, do not lead to overclosure
of the universe.

As we are going to consider rather small NC mass scales, it is no
longer relevant to restrain ourselves to the QCD axion case. We
will therefore keep only the model-independent bound (\ref{star
primakoff constraint on M}) on $M$. Similarly, the usual 'axion
window' for $m$ mentioned above should be rederived in the new
framework and will not be taken into account here.

\subsection{Oscillation length, ellipticity and
polarization rotation}

Since $|\omega^\pm_\| -\omega|\ll \omega$ and
$|\omega_\bot-\omega|\ll \omega$, one can approximate
(\ref{omega_parallel}) with
\begin{eqnarray}
\omega_\|^\pm &=& \omega_\bot
+\frac{1}{2}\Big(\frac{m^2}{2\omega}+g\vec{\theta}\cdot\vec{\bf
B}\omega  \Big)\pm \frac{1}{2}
\sqrt{(\frac{m^2}{2\omega}+g\vec{\theta}\cdot \vec{\bf B}\omega
)^2+\frac{\vec{\bf B}^2}{M^2}}
\nonumber \\
&=& \omega_\bot
+\frac{1}{2}\Big(\frac{m^2}{2\omega}+g\vec{\theta}\cdot\vec{\bf
B}\omega   \Big)\Big(1\pm \frac{1}{\cos(2\alpha)}\Big)
\end{eqnarray}
where
\begin{equation}
\alpha = \frac{1}{2}\arctan\Big(\frac{|\vec{\bf
B}|/M}{|\frac{m^2}{2\omega}+g\vec{\theta}\cdot\vec{\bf B}\omega
 |}\Big) \in
[0,\frac{\pi}{4}]
\end{equation}
is the photon-axion mixing angle. The weak mixing condition
$\alpha \ll 1$, or
\begin{equation}
\frac{m^2}{2\omega}+g\vec{\theta}\cdot \vec{\bf B}\omega
 \gg \frac{|\vec{\bf
B}|}{M} , \label{weak mixing condition}
\end{equation}
translates, in the usual case of vanishing or subdominant
non-commutativity compared to the axion mass, into a constraint on
the axion parameters,
\begin{equation}
m^2 M \gg |\vec{\bf B}| \omega  .
\end{equation}
However, we are interested in the case of {\it dominant}
non-commutativity in the left-hand side of (\ref{weak mixing
condition}), which is related to the oscillation length $L_{\rm
osc}$ by
\begin{equation}
2\pi L^{-1}_{\rm osc} =\frac{m^2}{2\omega}+g\vec{\theta}\cdot
\vec{\bf B}\omega \equiv 2\pi L^{-1}_{\rm axion} + 2\pi
L^{-1}_{\theta} . \label{inverse oscillation length}
\end{equation}
Assuming $\theta$ as large as allowed by the tree-level NC QED
bound $\theta \leq (10\ TeV)^{-2}$, as discussed in the
introduction, the $\theta$ term dominates the inverse oscillation
length for $m \leq \omega \sqrt{2g \vec{\theta}\cdot \vec{\bf
B}}\sim 10^{-12}\ eV$, a nearly massless axion, and the weak
mixing condition implies then an almost Planck-size axion coupling
$gM \geq 10^{17}\ GeV$. Although this is extremely small, note
that the QED contribution to the inverse oscillation length, that
should have been included in principle in (\ref{inverse
oscillation length}), is still subdominant,
\begin{equation}
2\pi L^{-1}_{QED}\sim \frac{2\alpha_{QED}^2|\vec{\bf
B}|^2\omega}{15m_e^4} < 2\pi L^{-1}_{\theta} \sim
g|\vec{\theta}\cdot \vec{\bf B}|\omega
\end{equation}
(with $m_e$ the electron mass) for a background magnetic field
$|\vec{\bf B}|$ of order $0.1\ T$ or lower. The Cotton-Mouton
contribution $L^{-1}_{CM}$ to the inverse oscillation length, due
to the residual gas in the cavity, has to be even smaller in any
experiment designed to observe the QED contributions, which can be
achieved for small enough gas pressure.

Under all these conditions, one should observe a rotation $\psi$
of the polarization plane after N traversals of a cavity of length
$L$, for an initial angle at 45$^\circ$ of the magnetic field,
\begin{equation}
\psi {\rm\ [rad]\ } \approx \frac{1}{8}\frac{N L^2|\vec{\bf
B}|^2}{M^2}\ ,
\end{equation}
which is independent of the oscillation length and an ellipticity
(ratio of minor to major axis)
\begin{equation}
{\cal E} \approx \frac{\pi}{12}\frac{N L^2 |\vec{\bf{B}}|^2}{M^2}
\ \frac{L}{L_{\rm osc}} \approx \frac{1}{24}\frac{N L^3
|\vec{\bf{B}}|^2}{M^2} g\vec{\theta}\cdot\vec{\bf B}\omega.
\end{equation}
The expansions above are allowed by the fact that we are largely
in the coherence regime $L \ll L_{\rm osc}$ necessary for the
experiment. The rotation angle $\psi$ is independent of the
oscillation length and is thus the same as in the commutative
axion case. It originates from a depletion of the photon component
parallel to the background field as it transforms into axions,
while the orthogonal component remains the same. It is
equivalently a measure of the photon-axion transition rate
depending essentially on $M$, which is exactly what the CAST
experiment, the star cooling rates, or the less constraining
$\gamma \rightarrow \phi \rightarrow \gamma$ 'shine light through
walls' \cite{Ruoso:1992nx} experiments measure. The polarization
rotation $\psi$ is thus not the most relevant number in PVLAS-type
oscillation experiments: only the induced ellipticity ${\cal E}$
is sensitive to the oscillation length, and thus can provide more
complete information on the detailed physics involved such as the
axion mass $m$ in the commutative case, or the NC scale
$\theta^{-1/2}$ here. Incidentally, this shows that known lower
bounds on $M$ should be unchanged by non-commutativity.

Unfortunately here, the weak mixing condition pushes $M$ to almost
Planckian values, and as we required also a practically massless
axion, the effect of mixing should be roughly the same as that of
photon-graviton mixing, unobservably small. Practically, $N \sim
10^5$ and $L=1\ m$, so that we expect for weak mixing
\begin{eqnarray}
\psi_{\rm weak} &\sim&
10^{-32}\Big(\frac{M_P}{M}\Big)^2\Big(\frac{|\vec{\bf{B}}|}{5.5\
T}\Big)^2 \ll (1-4)\times 10^{-7}\ {\rm rad}
\nonumber \\
{\cal E}_{\rm weak} &\sim& 10^{-49}\Big(\frac{M_P}{M}\Big)^2
\Big(\frac{|\vec{\bf{B}}|}{5.5\ T}\Big)^3 \ll 10^{-8}-2\times
10^{-7}
\end{eqnarray}
where the last numbers given are the ones measured by PVLAS. Also,
the ellipticity does not have the observed $\vec{\bf B}^2$
dependence in the background field. Finally, we note that the
experiment magnet is actually on a slowly rotating turntable such
that the magnetic field rotates in the horizontal plane, the beam
propagating vertically. The measured ellipticity would fluctuate
in phase with the magnetic field as the $\vec{\theta}\cdot
\vec{\bf B}$ term varies.

We may also consider the case of maximal mixing
\begin{equation}
\frac{m^2}{2\omega}+g\vec{\theta}\cdot\vec{\bf B}\omega \ll
\frac{\vec{|\bf B}|}{M} .
\end{equation}
In this case we do not expect the NC scale to play any role.
Effectively, no ellipticity is generated, and the only effect is a
polarization rotation, again unsensitive to the oscillation length
and identical to the commutative axion case.

\section{Conclusions}

We have studied noncommutative electromagnetic wave propagation in
a constant background field, and shown that contrary to previous
claims, no polarization rotation occurs for general
$\theta^{\mu\nu}$. We illustrated the practical importance of
using the commutative field variables derived from the
Seiberg-Witten map. The result has been cross-checked using the
general Lorentz-violating extension of QED. We have then studied
how axion-photon oscillations in a strong background field may
probe noncommutativity of space-time. Processes which are only
sensitive to the axion-photon coupling $M$ are unaffected, while
for small enough axion mass, the inverse oscillation length may be
dominated by the NC scale. PVLAS-type experiments measuring
induced ellipticity on linearly polarized light propagating in a
strong magnetic field can directly probe this scale in principle.
We found however that the almost Planckian axion-photon coupling
required, even for a very fine-tuned NC scale $\theta^{-1/2}\sim
10 \ TeV$, makes the effect unobservably small.
\\
\
\\

{\large \bf Acknowledgments}
\\

NC is supported by the NSF under grant PHY-0354990 and by Research
Corporation. NC and AP are supported by DOE grant
DE-FG02-85ER40231.

\end{document}